\documentclass[11pt,a4paper]{article}
\usepackage{jcappub}

\begin{document}

\title{Probing Cosmic Acceleration by Using the SNLS3 SNIa Dataset}

\author[a,b,c]{Xiao-Dong Li,}
\author[c,d]{Song Li,}
\author[a,c]{Shuang Wang,}
\author[a,c]{Wen-Shuai Zhang,}
\author[c,d,e]{Qing-Guo Huang}
\author[c,d,e]{and Miao Li}

\affiliation[a]{Department of Modern Physics, University of Science and Technology of China, Hefei 230026, China}
\affiliation[b]{Interdisciplinary Center for Theoretical Study, University of Science and Technology of China, Hefei 230026, China}
\affiliation[c]{Institute of Theoretical Physics, Chinese Academy of Science, Beijing 100080, China}
\affiliation[d]{Kavli Institute for Theoretical Physics China, Chinese Academy of Sciences, Beijing 100080, China}
\affiliation[e]{State Key Laboratory of Frontiers in Theoretical Physics, Chinese Academy of Sciences, Beijing 100190, China}

\emailAdd{renzhe@mail.ustc.edu.cn}
\emailAdd{sli@itp.ac.cn}
\emailAdd{swang@mail.ustc.edu.cn}
\emailAdd{wszhang@mail.ustc.edu.cn}
\emailAdd{huangqg@itp.ac.cn}
\emailAdd{mli@itp.ac.cn}

\abstract{
We probe the cosmic acceleration by using the recently released SNLS3 sample of 472 type Ia supernovae.
Combining this type Ia supernovae dataset with the cosmic microwave background anisotropy data from the Wilkinson Microwave Anisotropy Probe 7-yr observations,
the baryon acoustic oscillation results from the Sloan Digital Sky Survey data release 7,
and the Hubble constant measurement from the Wide Field Camera 3 on the Hubble Space Telescope,
we measure the dark energy equation of state $w$ and the deceleration parameter $q$ as functions of redshift by using the Chevallier-Polarski-Linder parametrization.
Our result is consistent with a cosmological constant at 1$\sigma$ confidence level,
without evidence for the recent slowing down of the cosmic acceleration.
Furthermore, we consider three binned parametrizations ($w$ is piecewise constant in redshift $z$) based on different binning methods.
The similar results are obtained, i.e., the $\Lambda$CDM model is still nicely compatible with current observations.
}


\maketitle

\section{Introduction}\label{sec:intro}

The cosmic acceleration has become one of the most important problems in modern cosmology since its discovery in 1998 \cite{Riess}.
This implies that around two thirds of components in our universe is composed of a mysterious dark energy (DE) \cite{DEReview}.
Although numerous theoretical DE models have been proposed in the past decade
\cite{quint}\cite{phantom}\cite{k}\cite{Chaplygin}\cite{tachyonic}\cite{refHDE}\cite{refADE}\cite{refRDE}\cite{hessence}\cite{YMC}\cite{Onemli},
the nature of DE still remains a mystery. Nowadays
an important question concerning the cosmic acceleration is to determine whether DE is consistent with a cosmological constant \cite{Weinberg}\cite{XZhang}.

A powerful probe of DE is Type Ia supernovae (SNIa),
which can be used as cosmological standard candles to directly measure the cosmic expansion.
Recently, a high-quality joint sample of 472 supernovae (SN), the SNLS3 SNIa dataset \cite{SNLS3A}, was released.
This SNIa sample includes 242 SN at $0.08 < z < 1.06$ from the Supernova Legacy Survey (SNLS) 3-yr observations \cite{SNLS3},
123 SN at low redshifts \cite{lowzSNIa}\cite{Constitution},
93 SN at intermediate redshifts from the Sloan Digital Sky Survey (SDSS)-II SN search \cite{SDSS-II},
and 14 SN at $z > 0.8$ from Hubble Space Telescope (HST) \cite{HSTSNIa}.
This SNIa sample has been used to study the evolution of DE equation of state (EOS) $w$ by Sullivan {\it et al.} \cite{SNLS3B}.
However, some other important quantities,
such as the DE density $\rho_{de}$ \cite{recrhode} and the deceleration parameter $q$ \cite{recqz}, have not been studied in their work.

In this work, we explore the cosmological consequences of the SNLS3 SNIa dataset.
In addition to the DE EOS $w$, we also study the evolution of deceleration parameter $q$.
To perform a comprehensive analysis including multiple observational techniques,
we combine the cosmic microwave background (CMB) anisotropy data from the Wilkinson Microwave Anisotropy Probe 7-yr (WMAP7) observations \cite{WMAP7},
the baryon acoustic oscillation (BAO) results from the SDSS Data Release 7 (DR7) \cite{SDSSDR7},
and the Hubble constant measurement from the Wide Field Camera 3 (WFC3) on the HST \cite{HSTWFC3}.

In 2009, by studying the popular Chevallier-Polarski-Linder (CPL) parametrization \cite{CPL},
Shafiello {\it et al.} \cite{Shafiello} argued that the Constitution SNIa sample \cite{Constitution} appears to support a dynamical DE.
Moreover, their analysis implies a possibility: the cosmic acceleration has already peaked, and we are currently witnessing its slowing down.
It is interesting to test whether this strange result is caused by the Constitution SNIa dataset itself.
So in this work we revisit the CPL parametrization by using the SNLS3 SNIa dataset.

Moreover, to further explore the cosmic acceleration,
we also take into account the binned parametrization.
This parametrization was firstly proposed by Huterer and Starkman \cite{BinnedPCA}
based on the principal component analysis (PCA) \cite{BinnedPCA}\cite{BinnedPCA2}.
The basic idea is to divide the redshift range into different bins and setting $w$ as piecewise constant in redshift $z$.
It should be mentioned that there are different methods for the optimal choice of the redshift bins in the literatures.
In \cite{Binned1Wang}, Wang argued that one should choose a constant $\Delta z$ for redshift slices.
This is because that the observables like $H$ and $1/D_A$ (length scales extracted from data analysis)
are assumed to be constant in each redshift slice in the galaxy redshift survey.
In \cite{Binned2Riess}, Riess {\it et al.} proposed another binning method, which has also drawn a lot of attention \cite{Binned2Works}.
In this method, the number of SNIa in each bin times the width of each bin is a constant (i.e. $n\Delta z =$ const).
In addition, in \cite{Binned3Li}, we presented a new binned parametrization method.
Instead of choosing the discontinuity points $z_i$ by hand, one can treat $z_i$ as models parameters and let them run freely.
Since all these choices are reasonable,
in this work we perform a comprehensive analysis and use all the above methods.

This paper is organized as follows:
In Sec. \ref{Models}, we describe the phenomenological models considered here and the method of data analysis.
In Sec. \ref{Data}, we introduce the observational data and describe how they are included in our analysis.
In Sec. \ref{Results} we present the results obtained in this paper.
In the end, we give a short summary in Sec. \ref{Conclusion}.
In this work, we assume today's scale factor $a_{0}=1$, so the redshift $z=a^{-1}-1$;
the subscript ``0'' always indicates the present value of the corresponding quantity, and the unit with $c=\hbar=1$ is used.

\section{Models and Methodology}\label{Models}

Standard candles impose constraints on cosmological parameters
through a comparison between the luminosity distance from observations and that from theoretical models.
In a spatially flat Friedmann-Robertson-Walker (FRW) universe (the assumption of flatness is motivated by the inflation scenario),
the luminosity distance $d_L$ is given by
\begin{equation}
\label{eq:dl}
d_L(z)=\frac{1+z}{H_{0}}\int_0^z\frac{dz'}{E(z')},
\end{equation}
with
\begin{equation}
\label{eq:Ez}
E(z)\equiv H(z)/H_{0} =\left[\Omega_{r}(1+z)^4+\Omega_m(1+z)^3+(1-\Omega_{r}-\Omega_{m})f(z)\right]^{1/2}.
\end{equation}
Here $H(z)$ is the Hubble parameter, $H_{0}$ is the Hubble constant, $\Omega_{m}$ is the present fractional matter density,
and $\Omega_{r}$ is the present fractional radiation density, given by \cite{WMAP7},
\begin{equation}
\Omega_{r}=\Omega_{\gamma}(1+0.2271N_{eff}),\ \ \
\Omega_{\gamma}=2.469\times10^{-5}h^{-2},\ \ \ N_{eff}=3.04,
\end{equation}
where $\Omega_{\gamma}$ is the present fractional photon density,
$h$ is the reduced Hubble parameter,
and $N_{eff}$ is the effective number of neutrino species.
The DE density function $f(z)\equiv \rho_{de}(z)/\rho_{de}(0)$ is a key function,
because a DE parametrization scheme enters in $f(z)$.

First, we consider the CPL model, which assumes
\begin{equation}\label{CPL_w}
w(z) = w_0 + w_a \frac{z}{1+z},
\end{equation}
where $w_0$ and $w_a$ are constants.
The corresponding $f(z)$ is
\begin{equation}\label{CPL_f}
f(z)=(1+z)^{3(1+w_0+w_a)}\exp\left(-\frac{3w_a z}{1+z}\right).
\end{equation}

Then, we consider the binned parametrization.
For this case, $f(z)$ takes the form
\begin{equation}\label{BinnedRhoDE}
f(z_{n-1}<z \le z_n)=(1+z)^{3(1+w_n)}\prod_{i=1}^{n-1}(1+z_i)^{3(w_i-w_{i+1})},
\end{equation}
where $w_i$ is the EOS in the $i$th redshift bin defined by an upper boundary at $z_i$.
In this work, we consider the case of 3 bins, i.e., $n=3$.

As mentioned above, we consider three binned parametrizations based on three binning methods.
For the first binning method where $\Delta z$ = const, we choose
\begin{equation}
z_1 = 0.5, \ \ z_2 = 1.0.
\end{equation}
We will call it ``const $\Delta z$'' model hereafter.

For the second binning method where $n\Delta z$ = const, we choose
\begin{equation}
z_1 = 0.3, \ \ z_2 = 0.73.
\end{equation}
In this way we have $n\Delta z \sim 65$ at the SNIa reshift region $0 < z \leq 1.4$.
We will call it ``const $n\Delta z$'' model hereafter.

Finally, in the last binning method, we perform a best-fit analysis,
and find the following choice
\begin{equation}
z_1 = 0.2, \ \ z_2 = 1.31,
\end{equation}
can yield a minimal $\chi^2$.
In our analysis, we set the conditions $z_1 > 0.1,\ z_2 > z_1 + 0.1$ and $z_2 < 1.4$.
We will call it ``free $\Delta z$'' model hereafter.

In this work, we adopt the $\chi^2$ statistic to estimate the model parameters.
For a physical quantity $\xi$ with experimentally measured value $\xi_{obs}$,
standard deviation $\sigma_{\xi}$ and theoretically predicted value $\xi_{th}$,
the $\chi^2$ takes the form
\begin{equation}
\chi^2_{\xi}={(\xi_{obs}-\xi_{th})^2\over \sigma^2_{\xi}}.
\end{equation}
The total $\chi^2$ is the sum of all $\chi^2_{\xi}$s, i.e.
\begin{equation}
\chi^2=\sum_{\xi}\chi^2_{\xi}.
\end{equation}
One can determine the best-fit model parameters by minimizing the total $\chi^2$.
Moreover, by calculating $\Delta \chi^2 \equiv \chi^2-\chi^2_{\rm min}$,
one can determine the 1$\sigma$ and the 2$\sigma$ confidence level (CL) ranges of a specific model.
Statistically, for models with different $n_p$ (denoting the number of free model parameters),
the 1$\sigma$ and 2$\sigma$ CL correspond to different $\Delta \chi^2$.
In Table \ref{NpDchisq}, we list the relationship between $n_p$ and $\Delta \chi^2$ from $n_p=1$ to $n_p=9$.

In this work,
we determining the best-fit parameters and the 1$\sigma$ and 2$\sigma$ CL ranges by using the Monte Carlo Markov chain (MCMC) technique.
We modify the publicly available CosmoMC package \cite{COSMOMC} and generate $O(10^6)$ samples for each set of results presented in this paper.
We also verify the reliability and accuracy of the code by using the Mathematica program \cite{Wolfram}.

\begin{table}
\caption{Relationship between number of free model parameters $n_p$ and $\Delta \chi^2$.}
\begin{center}
\label{NpDchisq}
\begin{tabular}{|c|c|c|}
  \hline
  ~~~$n_p$~~~ & ~~~$\Delta \chi^2(1\sigma)$~~~ & ~~~$\Delta \chi^2(2\sigma)$~~~ \\
  \hline
  ~~~  1  ~~~ &          ~~~  $1$  ~~~        &          ~~~  $4$  ~~~          \\
  \hline
  ~~~  2  ~~~ &          ~~~$2.30$ ~~~        &          ~~~$6.18$ ~~~          \\
  \hline
  ~~~  3  ~~~ &          ~~~$3.53$ ~~~        &          ~~~$8.02$ ~~~          \\
  \hline
  ~~~  4  ~~~ &          ~~~$4.72$ ~~~        &          ~~~$9.72$ ~~~          \\
  \hline
  ~~~  5  ~~~ &          ~~~$5.89$ ~~~        &          ~~~$11.31$ ~~~         \\
  \hline
  ~~~  6  ~~~ &          ~~~$7.04$ ~~~        &          ~~~$12.85$ ~~~         \\
  \hline
  ~~~  7  ~~~ &          ~~~$8.18$ ~~~        &          ~~~$14.34$ ~~~         \\
  \hline
  ~~~  8  ~~~ &          ~~~$9.30$ ~~~        &          ~~~$15.79$ ~~~         \\
  \hline
  ~~~  9  ~~~ &          ~~~$10.42$ ~~~        &          ~~~$17.21$ ~~~         \\
  \hline
\end{tabular}
\end{center}
\end{table}

\section{Observational data}\label{Data}

In this paper, we use the SNLS3 SNIa sample \cite{SNLS3A},
the CMB anisotropy data from the WMAP7 observations \cite{WMAP7},
the BAO results from the SDSS DR7 \cite{SDSSDR7},
and the Hubble constant measurement from the WFC3 on the HST \cite{HSTWFC3}.
In the following, we briefly describe how these data are included into the $\chi^2$ analysis.

\subsection{The SNIa data}

Here we use the SNLS3 SNIa dataset released in \cite{SNLS3A}.
This combined sample consists of 472 SN at $0.01 < z < 1.4$,
including 242 SN over $0.08 < z < 1.06$ from SNLS 3-yr observations \cite{SNLS3},
123 SN at low redshifts \cite{lowzSNIa}\cite{Constitution}, 93 SN at intermediate redshifts from the SDSS-II SN search \cite{SDSS-II},
and 14 SN at $z > 0.8$ from HST \cite{HSTSNIa}.
The systematic uncertainties of the SNIa data were nicely handled \cite{SNLS3A}.
The total data of the SNLS3 sample can be downloaded from \cite{SNLS3Code}.

The $\chi^2$ of the SNIa data is
\begin{equation}
\chi^2_{\rm SN}=\Delta \overrightarrow{\bf m}^T \cdot {\bf C}^{-1} \cdot \Delta \overrightarrow{\bf m},
\end{equation}
where {\bf C} is a $472 \times 472$ covariance matrix capturing the statistic and systematic uncertainties of the SNIa sample,
and $\Delta {\overrightarrow {\bf m}} = {\overrightarrow {\bf m}}_B - {\overrightarrow {\bf m}}_{\rm mod}$ is a vector of model residuals of the SNIa sample.
Here $m_B$ is the rest-frame peak $B$ band magnitude of the SNIa,
and $m_{\rm mod}$ is the predicted magnitude of the SNIa given by the cosmological model
and two other quantities (stretch and color) describing the light-curve of the particular SNIa.
The model magnitude $m_{\rm mod}$ is given by
\begin{equation}\label{SNchisq}
m_{\rm mod} = 5\log_{10} \mathcal{D}_L(z_{\rm hel},z_{\rm cmb})-\alpha(s-1)+\beta \mathcal{C} + \mathcal{M}.
\end{equation}
Here $\mathcal{D}_L$ is the Hubble-constant free luminosity distance, which takes the form
\begin{equation}
\mathcal{D}_L(z_{\rm hel}, z_{\rm zcmb}) = (1+z_{\rm hel})\int^{z_{\rm cmb}}_0 {{d z^\prime}\over{E(z^\prime)}},
\end{equation}
where $z_{\rm cmb}$ and $z_{\rm hel}$ are the CMB frame and heliocentric redshifts of the SN,
$s$ is the stretch measure for the SN,
and $\mathcal{C}$ is the color measure for the SN.
$\alpha$ and $\beta$ are nuisance parameters which characterize the
stretch-luminosity and color-luminosity relationships, respectively.
Following \cite{SNLS3A}, we treat $\alpha$ and $\beta$ as free parameters and let them run freely.

The quantity $\mathcal{M}$ in Eq. (\ref{SNchisq}) is a nuisance parameter
representing some combination of the absolute magnitude of a fiducial SNIa and the Hubble constant.
In this work, we marginalize $\mathcal{M}$ following the complicated formula in the Appendix C of \cite{SNLS3A}.
This procedure includes the host-galaxy information \cite{SullivanHostGalaxy} in the cosmological fits
by splitting the samples into two parts and allowing the absolute magnitude to be different between these two parts.

The total covariance matrix {\bf C} in Eq. (\ref{SNchisq}) captures both the statistical and systematic uncertainties of the SNIa data.
One can decompose it as \cite{SNLS3A},
\begin{equation}
{\bf C} = {\bf D}_{\rm stat} + {\bf C}_{\rm stat} + {\bf C}_{\rm sys},
\end{equation}
where ${\bf D}_{stat}$ is the purely diagonal part of the statistical uncertainties,
${\bf C}_{\rm stat}$ is the off-diagonal part of the statistical uncertainties,
and ${\bf C}_{\rm sys}$ is the part capturing the systematic uncertainties.
It should be mentioned that, for different $\alpha$ and $\beta$, these covariance matrices are also different.
Therefore, in practice one has to reconstruct the covariance matrix $\bf C$ for the corresponding values of $\alpha$ and $\beta$,
and calculate its inversion.
For simplicity, we do not describe these covariance matrices one by one.
One can refer to the original paper \cite{SNLS3A} and the public code \cite{SNLS3Code}
for more details about the explicit forms of the covariance matrices and the details of the calculation of $\chi^2_{\rm SN}$.

\subsection{The CMB data}

Here we use the ``WMAP distance priors'' given by the 7-yr WMAP observations \cite{WMAP7}.
The distance priors include the ``acoustic scale'' $l_A$, the ``shift parameter'' $R$, and the redshift of the decoupling epoch of photons $z_*$.
The acoustic scale $l_A$, which represents the CMB multipole corresponding to the location of the acoustic peak,
is defined as \cite{WMAP7}
\begin{equation}
\label{ladefeq} l_A\equiv (1+z_*){\pi D_A(z_*)\over r_s(z_*)}.
\end{equation}
Here $D_A(z)$ is the proper angular diameter distance, given by
\begin{equation}
D_A(z)=\frac{1}{1+z}\int^z_0\frac{dz^\prime}{H(z^\prime)},
\label{eq:da}
\end{equation}
and $r_s(z)$ is the comoving sound horizon size, given by
\begin{equation}
r_s(z)=\frac{1} {\sqrt{3}}  \int_0^{1/(1+z)}  \frac{ da } { a^2H(a)
\sqrt{1+(3\Omega_{b}/4\Omega_{\gamma})a} },
\label{eq:rs}
\end{equation}
where $\Omega_{b}$ and $\Omega_{\gamma}$ are the present baryon and photon density parameters, respectively.
In this paper, we adopt the best-fit values, $\Omega_{b}=0.02253 h^{-2}$ and $\Omega_{\gamma}=2.469\times10^{-5}h^{-2}$ (for $T_{cmb}=2.725$ K),
given by the 7-yr WMAP observations \cite{WMAP7}.
The fitting function of $z_*$ was proposed by Hu and Sugiyama \cite{Hu:1995en}:
\begin{equation}
\label{zstareq} z_*=1048[1+0.00124(\Omega_b
h^2)^{-0.738}][1+g_1(\Omega_m h^2)^{g_2}],
\end{equation}
where
\begin{equation}
g_1=\frac{0.0783(\Omega_b h^2)^{-0.238}}{1+39.5(\Omega_b h^2)^{0.763}},
\quad g_2=\frac{0.560}{1+21.1(\Omega_b h^2)^{1.81}}.
\end{equation}
In addition, the shift parameter $R$ is defined as \cite{Bond97}
\begin{equation}
\label{shift} R(z_*)\equiv \sqrt{\Omega_m H_0^2}(1+z_*)D_A(z_*).
\end{equation}
This parameter has been widely used to constrain various cosmological models \cite{Add3}.

As shown in \cite{WMAP7}, the $\chi^2$ of the CMB data is
\begin{equation}
\chi_{CMB}^2=(x^{obs}_i-x^{th}_i)(C_{CMB}^{-1})_{ij}(x^{obs}_j-x^{th}_j),\label{chicmb}
\end{equation}
where $x_i=(l_A, R, z_*)$ is a vector,
and $(C_{CMB}^{-1})_{ij}$ is the inverse covariance matrix.
The 7-yr WMAP observations \cite{WMAP7} had given the maximum likelihood values:
$l_A(z_*)=302.09$, $R(z_*)=1.725$, and $z_*=1091.3$.
The inverse covariance matrix was also given in \cite{WMAP7}
\begin{equation}
(C_{CMB}^{-1})=\left(
  \begin{array}{ccc}
    2.305 & 29.698 & -1.333 \\
    29.698 & 6825.27 & -113.180 \\
    -1.333 & -113.180  &  3.414 \\
  \end{array}
\right).
\end{equation}

\subsection{The BAO data}

Here we use the distance measures from the SDSS DR7 \cite{SDSSDR7}.
One effective distance measure is the $D_V(z)$, which can be obtained from the spherical average \cite{Eisenstein}
\begin{equation}
 D_V(z) \equiv \left[(1+z)^2D_A^2(z)\frac{z}{H(z)}\right]^{1/3},
\end{equation}
where $D_A(z)$ is the proper angular diameter distance.
In this work we use two quantities $d_{0.2}\equiv r_s(z_d)/D_V(0.2)$ and $d_{0.35}\equiv r_s(z_d)/D_V(0.35)$.
The expression of $r_s$ is given in Eq.(\ref{eq:rs}),
and $z_d$ denotes the redshift of the drag epoch, whose fitting formula is proposed by Eisenstein and Hu \cite{BAODefzd}
\begin{equation}
\label{Defzd} z_d={1291(\Omega_mh^2)^{0.251}\over 1+0.659(\Omega_mh^2)^{0.828}}\left[1+b_1(\Omega_bh^2)^{b2}\right],
\end{equation}
where
\begin{eqnarray}\label{Defb1b2}
b_1 &=& 0.313(\Omega_mh^2)^{-0.419}\left[1+0.607(\Omega_mh^2)^{0.674}\right], \\
b_2 &=& 0.238(\Omega_mh^2)^{0.223}.
\end{eqnarray}
Following \cite{SDSSDR7}, we write the $\chi^2$ for the BAO data as,
\begin{equation}
\chi^2_{BAO}=\Delta{p_i}(C_{BAO}^{-1})_{ij}\Delta p_j,
\end{equation}
where
\begin{equation}
\Delta p_i = p^{\rm data}_i - p_i,
\ \  p^{\rm data}_1 = d^{\rm data}_{0.2 } = 0.1905,
\ \  p^{\rm data}_2 = d^{\rm data}_{0.35} = 0.1097,
\end{equation}
and the inverse covariance matrix takes the form
\begin{equation}
(C_{BAO}^{-1})=\left(
  \begin{array}{cc}
    30124  & -17227 \\
    -17227 & 86977 \\
  \end{array}
\right).
\end{equation}

\subsection{The Hubble constant data}

The precise measurements of $H_0$ will be helpful to break the degeneracy between it and the DE parameters \cite{H0Freedman}.
When combined with the CMB measurement, it can lead to precise mesure of the DE EOS $w$ \cite{H0WHu}.
Recently, using the WFC3 on the HST, Riess {\it et al.} obtained an accurate determination of the Hubble constant \cite{HSTWFC3}
\begin{equation}
H_0=73.8\pm 2.4 {\rm km/s/Mpc},
\end{equation}
corresponding to a $3.3\%$ uncertainty.
So the $\chi^2$ of the Hubble constant data is
\begin{equation}
\chi^2_{h}=\left({h-0.738\over 0.024}\right)^2.
\end{equation}

\subsection{The total $\chi^2$}

Since the SNIa, CMB, BAO and $H_0$ are effectively independent measurements,
we can combine them by simply adding together the $\chi^2$ functions,
i.e.,
\begin{equation}
\chi^2_{All} = \chi^2_{SN} + \chi^2_{CMB} + \chi^2_{BAO} + \chi^2_{h}.
\end{equation}

\section{Results}\label{Results}

As mentioned above, we study the CPL model and the three binned models (const $\Delta z$, const $n\Delta z$ and free $\Delta z$) in this work.
A brief summary of these models are shown in Table \ref{SumofResults},
where the models, model parameters (together with their best-fit values and 1$\sigma$ uncertainties), and $\chi^2_{min}$s are given.
The nuisance parameters $h$, $\alpha$ and $\beta$ used in the analysis are actually not model parameters with significant meanings,
so we do not list them in the table.

To make a comparison,
we also listed the value of $\chi^2_{min}$ for the $\Lambda$CDM model.
The result shows that the inclusion of extra parameters ($w_0$, $w_a$, $w_1$, $w_2$) does not lead to a remarkable reduction of the $\chi^2_{min}$.
This implies that the $\Lambda$CDM model can provide a nice fit to the current data,
and the consideration of more complex model is not preferred.
In fact, the $\Lambda$CDM model usually has the best performance under the model-independent consideration \cite{ModelComp},
based on the ``information criteria'' \cite{IC}.
In this work, since our main purpose is to explore the cosmological consequences of the SNLS3 sample and to probe the cosmic acceleration,
we will not discuss the topic of the model comparisons.

\begin{table}[!htb]
\footnotesize
\caption{Summary of the models considered}
\begin{center}
\label{SumofResults}
\begin{tabular}{lcccccc}
\hline
~~~Model & $\Omega_m$ & $w_0$ & $w_a$ & $w_1$ & $w_2$ & $\chi^2_{min}$ \\
\hline
$\Lambda$CDM & $0.265^{+0.014}_{-0.013}$ & -- & -- & -- & -- & 424.911 \\
\hline
CPL & $0.265^{+0.027}_{-0.024}$ & $-1.07^{+0.32}_{-0.29}$ & $-0.09^{+1.13}_{-1.99}$ & -- & -- & 423.432 \\
\hline
const $\Delta z$ & $0.273^{+0.03}_{-0.032}$  & $-1.11^{+0.18}_{-0.17}$ & -- & $-2.43^{+2.45}_{-0.92}$ & $-0.03^{+0.06}$$\ast$ & 422.513 \\
\hline
const $n\Delta z$ & $0.268^{+0.038}_{-0.027}$ & $-1.06^{+0.21}_{-0.21}$ & -- & $-1.71^{+0.96}_{-1.41}$ & $-0.41^{+0.43}_{-1.76}$ & 421.642 \\
\hline
free $\Delta z$ & $0.276^{+0.031}_{-0.029}$ & $-0.97^{+0.24}_{-0.25}$ & -- & $-1.55^{+0.54}_{-0.32}$ & $-0.02^{+0.06}_{-3.39}$ & 419.737 \\
\hline
\end{tabular}
\center{$\ast${ For the const $\Delta z$ model,
the lower bound of $w_2$ is only weakly constrained,
so we do not list its value.}}
\end{center}
\end{table}

\subsection{The results of the CPL model}

\begin{figure}
\begin{center}
\includegraphics[width=16cm]{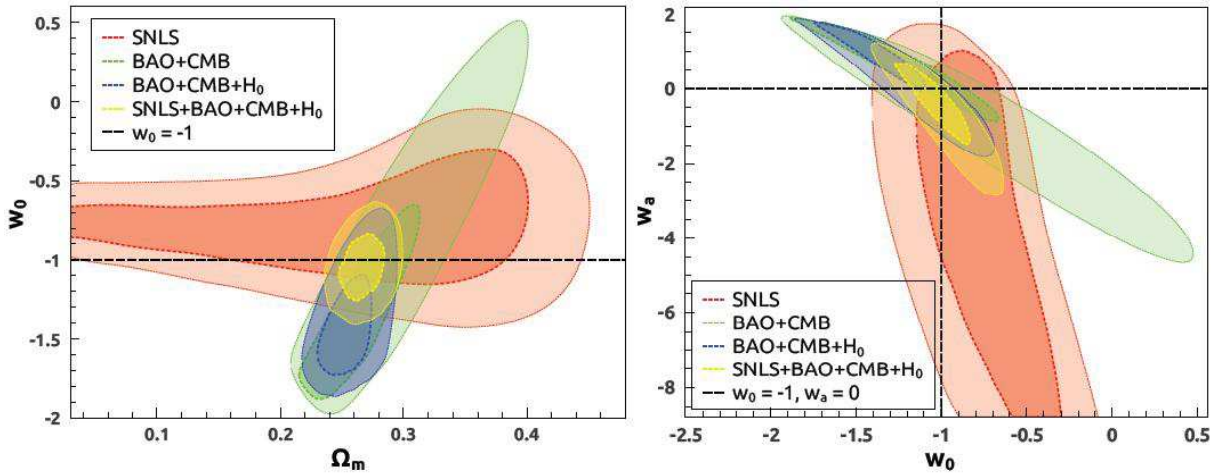}
\end{center}
\caption{\label{figCPLcontour} Marginalized probability contours at
1$\sigma$ and 2$\sigma$ CL in the $\Omega_m-w_0$ and $w_0-w_a$
planes, for the CPL model. Constraints from SNLS3, BAO+CMB,
BAO+CMB+$H_0$, and SNLS3+BAO+CMB+$H_0$ are all shown.
Our results are consistent with the Fig. 6 of \cite{SNLS3B}.
No evidence of deviations from the $\Lambda$CDM model is found.}
\end{figure}

\begin{figure}
\begin{center}
\includegraphics[width=7cm]{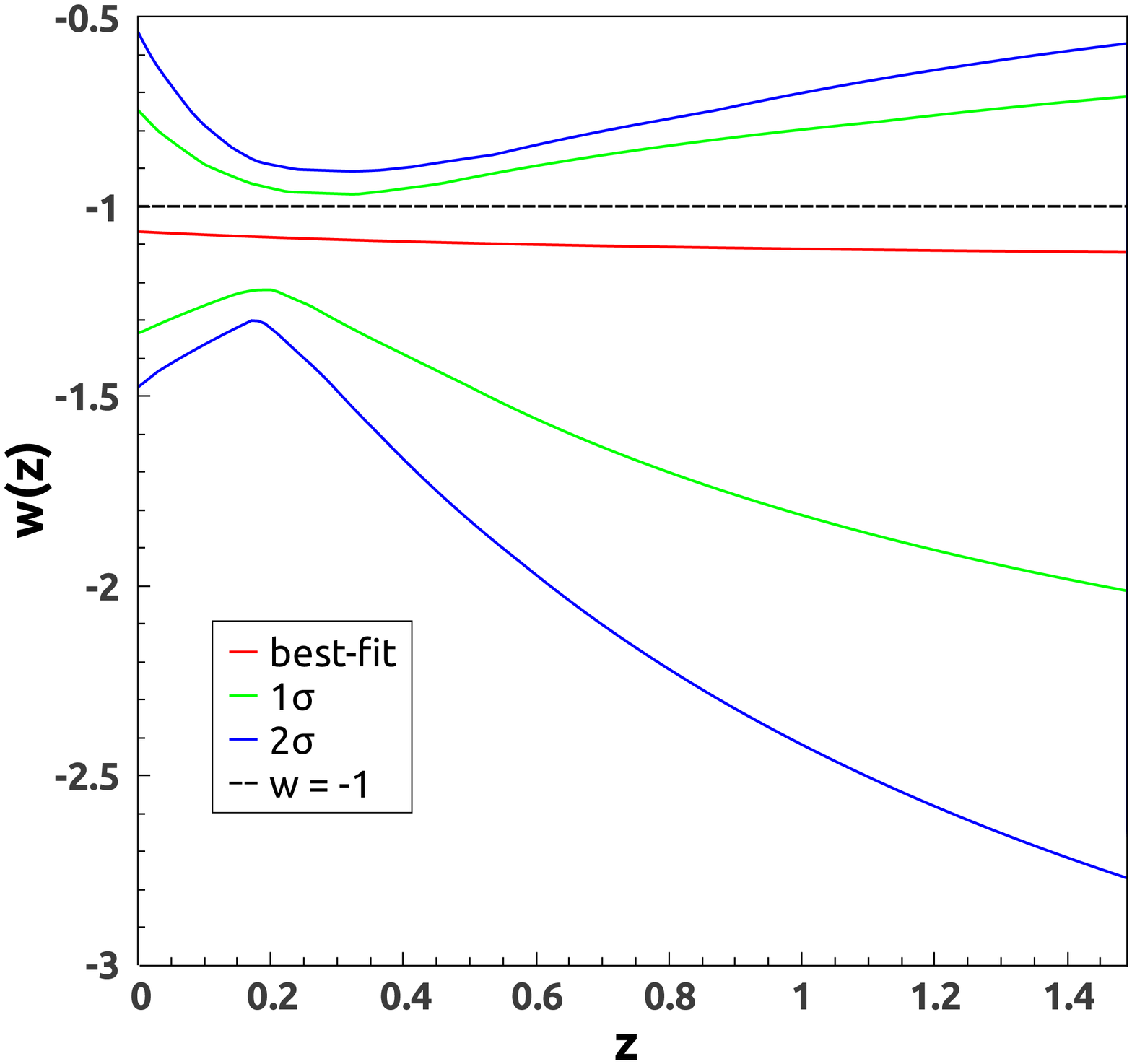}
\includegraphics[width=7cm]{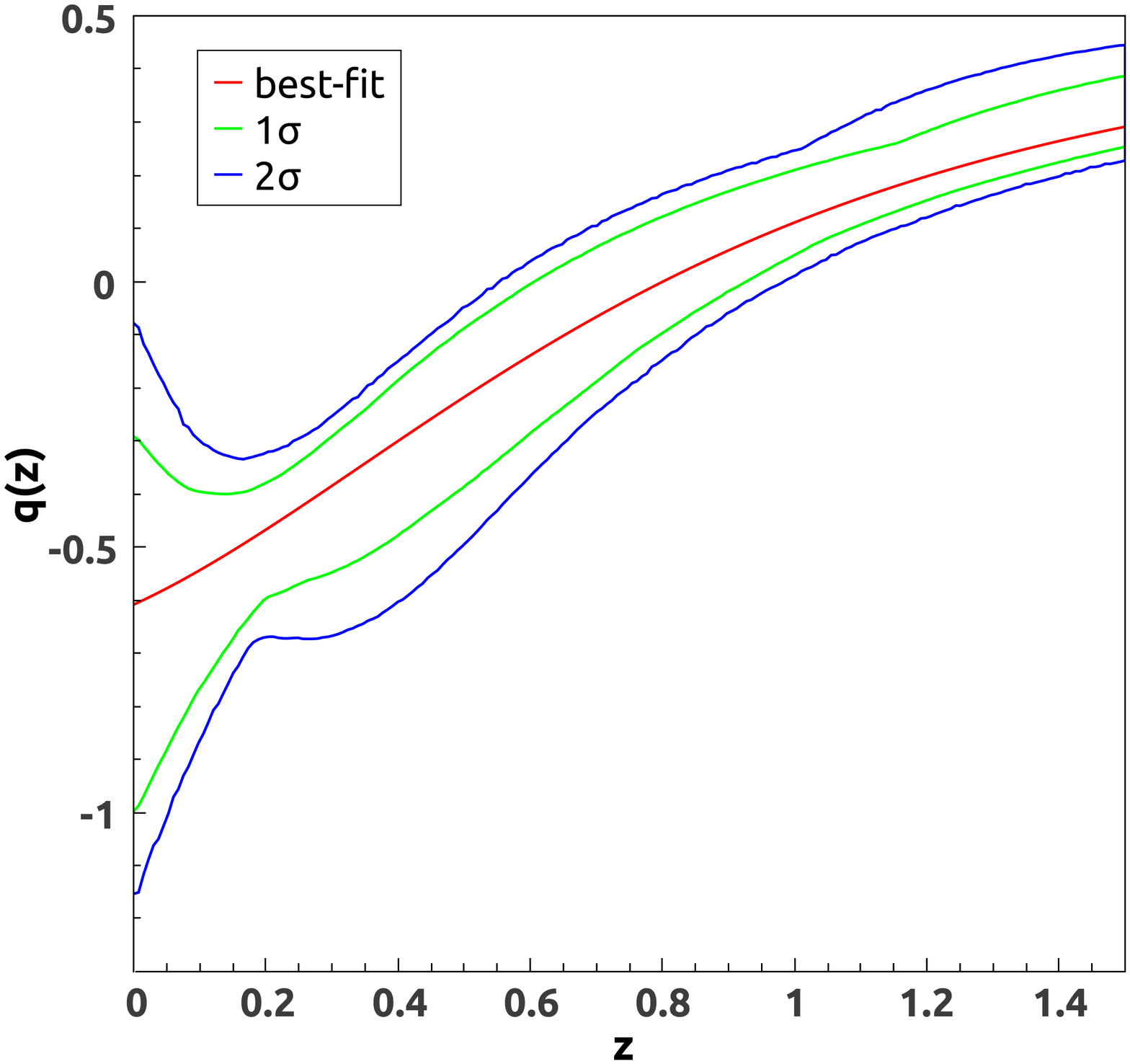}
\end{center}
\caption{\label{figCPLwq} The evolution of $w(z)$ and $q(z)$ along with z for the CPL model.
The result is consistent with the cosmological constant in the 1$\sigma$ CL.
Moreover, $q(z)$ is an increasing function of $z$, which is different from the result of \cite{Shafiello}.}
\end{figure}

The SNLS3 sample has been used to investigate the CPL model by
Sullivan et. al in \cite{SNLS3B}. By performing a global fit, they
obtained $w_0=-0.905^{+0.196}_{-0.196}$ and $w_a=-0.984^{+1.094}_{-1.097}$,
which is consistent with the $\Lambda$CDM model. In this work, we obtained the similar results.
In Fig. \ref{figCPLcontour}, we plot the 1$\sigma$ and 2$\sigma$ CL
contours of the CPL model in the $\Omega_m-w_0$ and $w_0-w_a$
planes, respectively. Constraints from SNLS3, BAO+CMB, BAO+CMB+$H_0$
and SNLS3+BAO+CMB+$H_0$ are shown in contours with differnt colors.
The overlaps of these contours imply that these cosmological
techniques lead to consistent cosmological implications in the CPL
ansatz. We find that the result is consistent with the $\Lambda$CDM
($w_0=-1$ and $w_a=0$) at the 1$\sigma$ CL. This is different from
the result of \cite{Shafiello}, where the authors found that the
$\Lambda$CDM appears to be in tension with the observations (see
Fig. 5 of \cite{Shafiello}).

To have a more direct view on the property of DE and the cosmic expansion,
in Fig. \ref{figCPLwq} we plot the evolution of the EOS $w(z)$ and the deceleration parameter $q(z)$ along with $z$.
From the result of $w(z)$, one can see that $\Lambda$CDM is within the 1$\sigma$ CL.
Moreover, the evolution of $q(z)$ shows that the current universe is under accelerating expansion.
No evidence of the recent slowing down of the cosmic acceleration is found.
This is also different from the result of \cite{Shafiello}.
The difference between the result of \cite{Shafiello} and ours may arise from the difference of the data used in the two papers.

\subsection{The results of the three binned models}

\begin{figure}
\begin{center}
\includegraphics[width=12cm,height=6cm]{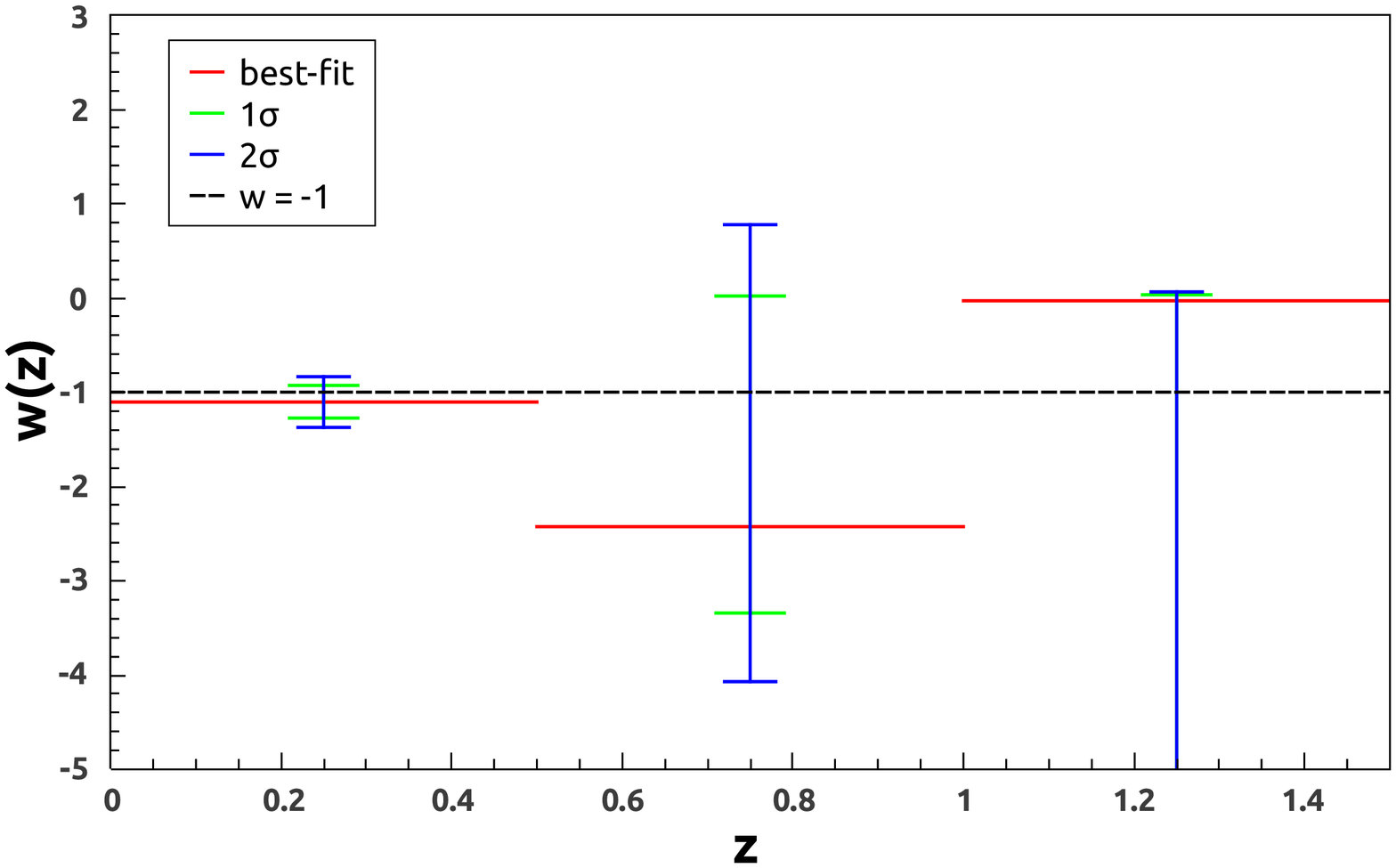}
\includegraphics[width=11cm,height=6cm]{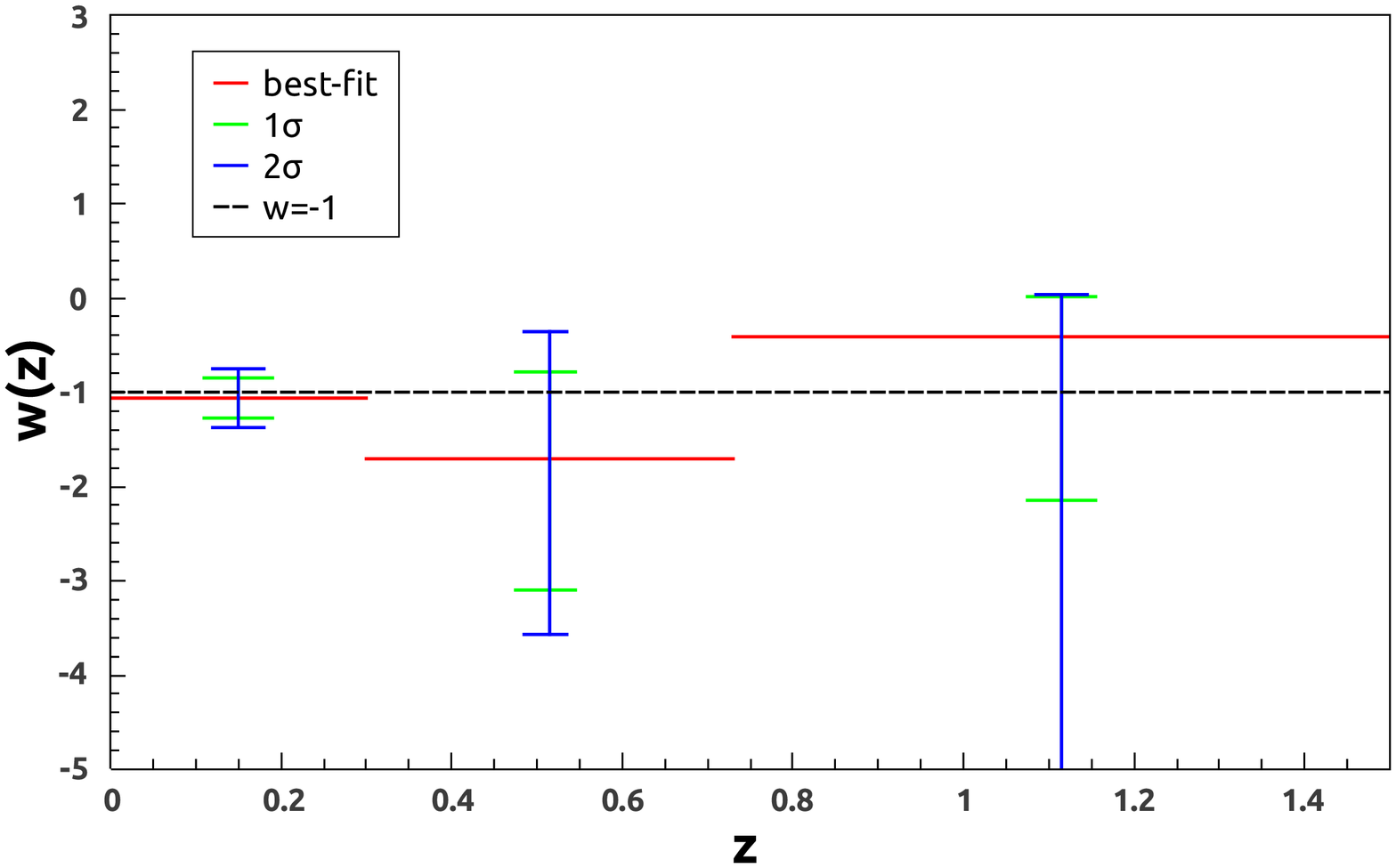}
\includegraphics[width=11cm,height=6cm]{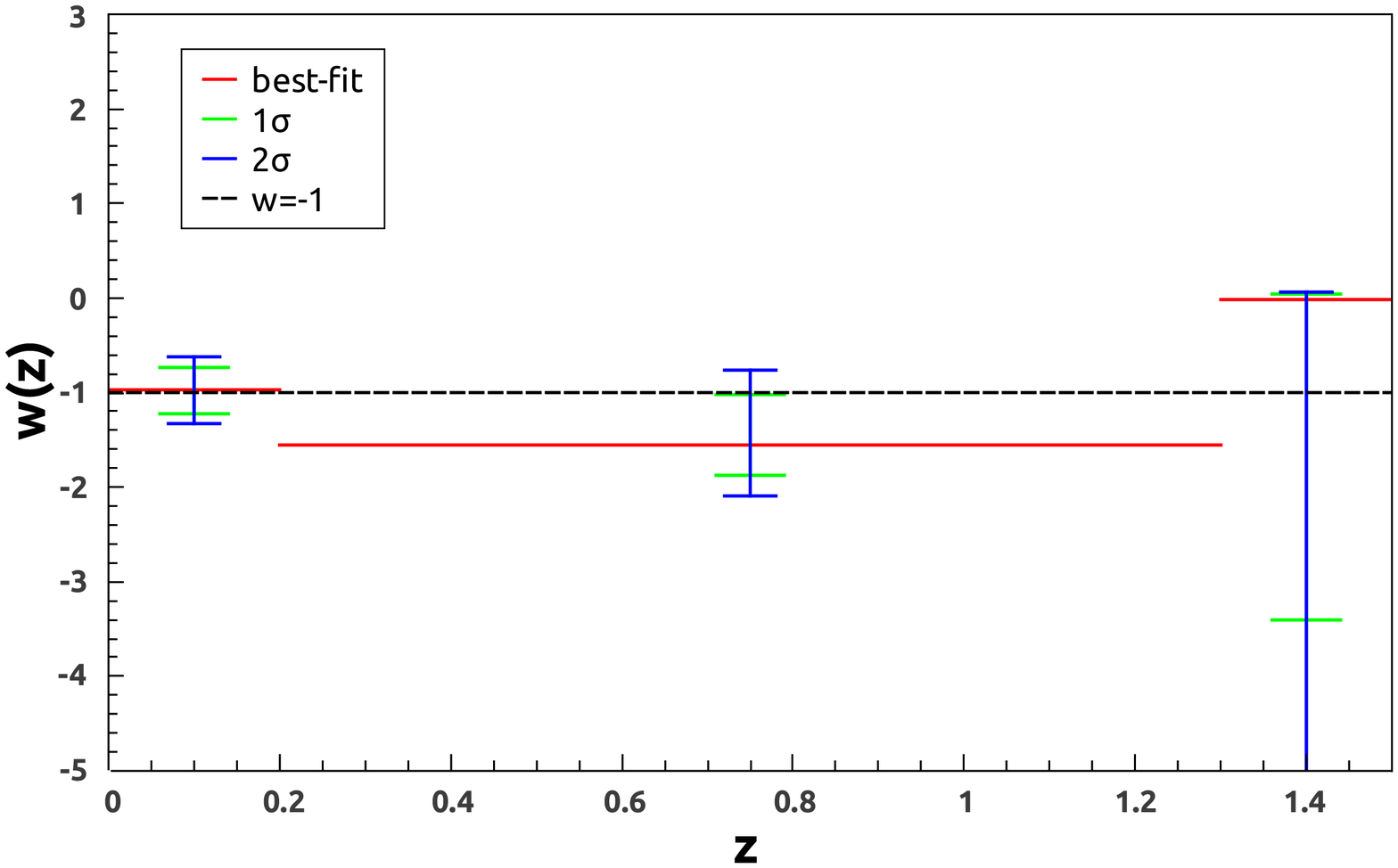}
\caption{\label{figBinnedwz} The evolution of $w(z)$ along with $z$ for the three binned models.
From top to bottom, the results of the const $\Delta z$, the const $n\Delta z$, and the free $\Delta z$ models are shown.
}
\end{center}
\end{figure}

In Fig. \ref{figBinnedwz}, we plot the evolution of $w(z)$ for the three binned models.
The best-fit as well as the 1$\sigma$ and 2$\sigma$ error bars are shown.
From Fig. \ref{figBinnedwz},
it is clear that the error of $w(z)$ in a certain redshift region depends on the number of SNIa in that region.
The larger the number of SNIa, the smaller the error is.
For example, the evolution of DE is most tightly constrained in the first redshift region
(about a half of the SNIa in the SNLS3 sample lie in the redshift region $z\leq0.3$),
and is only weakly constrained in the last region at $z \gtrsim 1.0$ (only $10-30$ SN lie in this region).
This signature also manifests itself in the result of the free $\Delta z$ model,
which has most tightly constrained $w(z)$ in the second redshift bin.

Except for  slight difference in the values of best-fit and errors,
the results of these three models are similar to each other,
with the following common features:

\noindent {\bf (i)} The results of the three binned models are all consistent with the $\Lambda$CDM in the 1$\sigma$ CL.

\noindent {\bf (ii)} For the last redshift region,
the upper bound of $w(z)$ is constrained to $w\lesssim 0.05$, mainly due to the inclusion of the CMB data.
The lower bound is only weakly constrained.

\noindent {\bf (iii)} At low redshifts, the result is tightly constrained and well consistent with the $\Lambda$CDM model.
So the evolution of $w(z)$ shows no evidence for the recent slowing down of the cosmic acceleration.

\begin{figure}
\includegraphics[width=16cm]{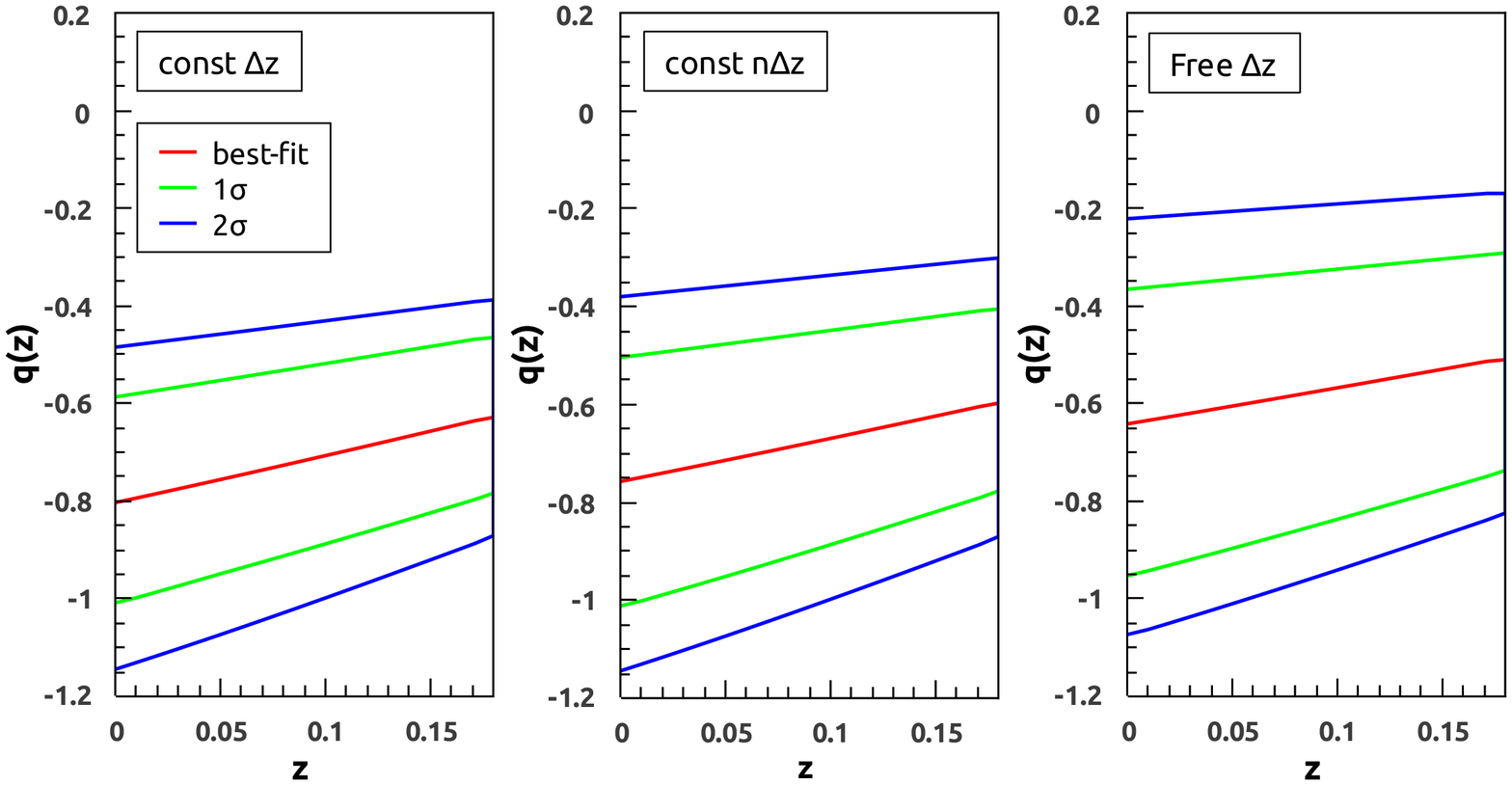}
\caption{\label{figBinnedqz} The evolution of $q(z)$ along with redshift for the three binned models at the low redshift.
From left to right, we show the results of the const $\Delta z$, the const $n\Delta z$, and the free $\Delta z$ models.
We found no evidence for the slowing down of the cosmic acceleration.
}
\end{figure}

To further confirm conclusion {\bf (iii)},
we also plot the evolution of $q(z)$ for the three models at the low redshifts ($z<0.2$) in Fig. \ref{figBinnedqz}.
The results of the three binned models all indicate a universe with continuous accelerated expansion.
This feature is in accordance with the result of the CPL model.

\begin{figure}
\includegraphics[width=16cm]{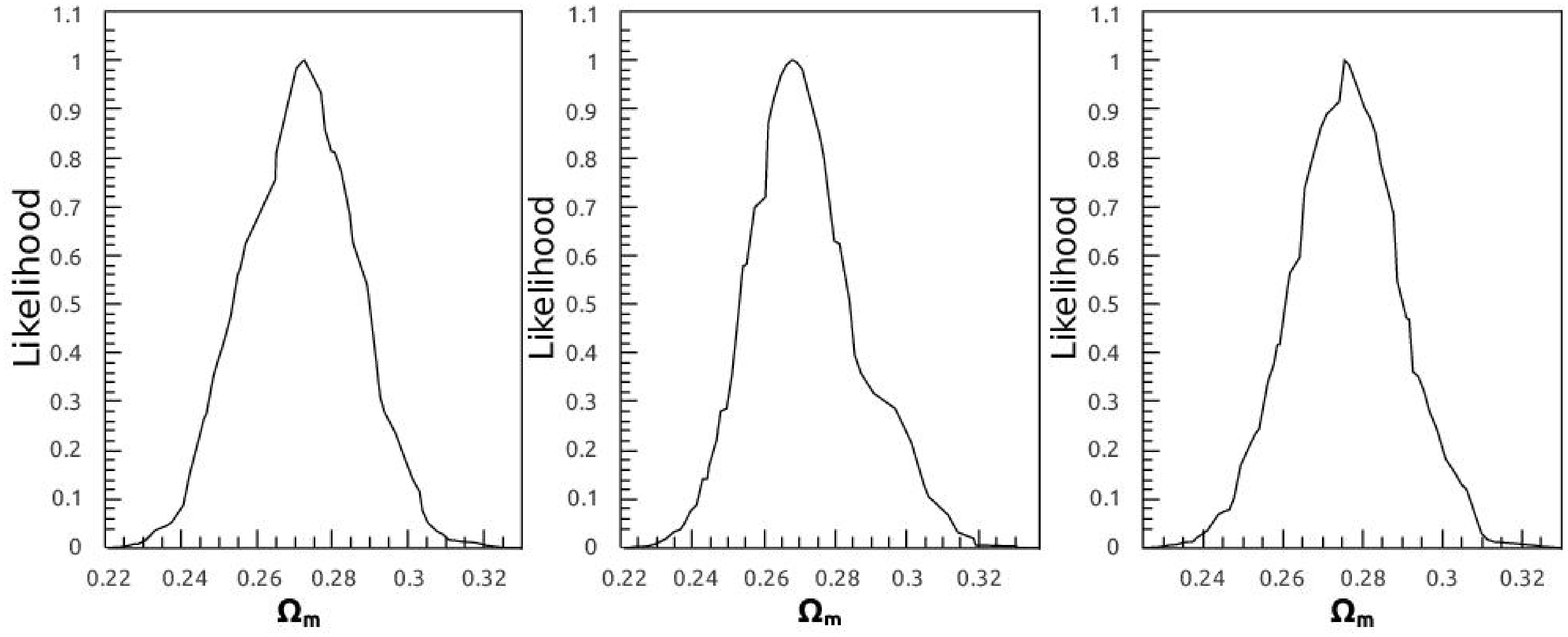}
\caption{\label{figBinnedOmegam} The distribution of the likelihood ${\mathcal L}\propto e^{-\chi^2/2}$ for the three binned models.
From left to right, the results of the const $\Delta z$, the const $n\Delta z$, and the free $\Delta z$ models are shown.}
\end{figure}

In addition, we also plot the constraints on $\Omega_m$ for the these three binned models in Fig. \ref{figBinnedOmegam}.
The figure shows that these three models lead to similar constraint on $\Omega_m$, with the best-fit value of 0.26-0.28,
which is also consistent with the constraint on $\Omega_m$ in the $\Lambda$CDM model and the CPL model (see Table II).
Therefore, we obtain consistent results from these four models.

\section{Conclusion}\label{Conclusion}

In this work we probe the cosmic acceleration by using various cosmological observations,
including the SNLS3 SNIa sample, the CMB data from the WMAP7 observations,
the BAO data from the SDSS DR7, and the Hubble constant measurement from the HST.
For models, we consider the CPL parametrization and three binned parameterizations based on different binning methods.
We focus on the reconstruction of the DE EOS $w(z)$ and the deceleration parameter $q(z)$,
to study whether the cosmic acceleration is slowing down, mentioned by Shafiello {\it et al.} in \cite{Shafiello}.
Adopting four models in this paper,
we obtain consistent results: the cosmological constant is still consistent with the data at 1$\sigma$ CL,
and there is no evidence for the slowing down of the cosmic acceleration.
Our result is different from the result of \cite{Shafiello},
which may arise from the difference of the data used in the two papers.
Although our result supports the standard $\Lambda$CDM paradigm,
the uncertainties in the reconstruction of $w(z)$ still allows space for other dynamical DE models.
So the current observational data are still too limited to determine whether the cosmic acceleration is driven by a cosmological constant.

\

\

\

\

\section*{Acknowledgements}
We would like to thank Alexander Conley for helpful discussions.
This work was supported by the NSFC grant No.10535060/A050207,
a NSFC group grant No.10821504 and Ministry of Science and Technology 973 program under grant No.2007CB815401.
QGH was also  supported by the project of Knowledge Innovation
Program of Chinese Academy of Science and a grant from NSFC (Grant No. 10975167).

\

\

\

\


\end{document}